\documentclass[12pt, aps,reprint, twocolumns, superscriptaddress,longbibliography]{revtex4-2}
\usepackage[a4paper, total={6in, 8in}]{geometry}
\usepackage{graphicx}
\usepackage{color}
\usepackage{upgreek}
\usepackage{graphics}
\usepackage{enumitem}
\usepackage{amsmath}
\usepackage{float}
\usepackage{siunitx}
\usepackage{amsfonts}
\usepackage[table]{xcolor} 
\usepackage{lineno} 

\usepackage{hyperref}
\hypersetup{
    colorlinks=true,
    linkcolor=blue,
    filecolor=magenta,      
    urlcolor=cyan,
    }
\usepackage{setspace}

\makeatletter
\newcommand{\maketitlepage}{%
  \begin{titlepage}
    \let\thanks\@gobble
    \let\footnote\@gobble
    \if@twocolumn
      \ifnum \col@number=\@ne
        \@maketitle
      \else
        \twocolumn[\@maketitle]%
      \fi
    \else
      \@maketitle
    \fi
    \thispagestyle{empty}
  \end{titlepage}%
}
\makeatother

\begin{document}


\title{Geometric deep learning reveals the spatiotemporal fingerprint of microscopic motion}

\author{Jes\'{u}s Pineda}
\affiliation{Department of Physics, University of Gothenburg, Origov{\"a}gen 6B, SE-41296 Gothenburg, Sweden}
\author{Benjamin Midtvedt}
\affiliation{Department of Physics, University of Gothenburg, Origov{\"a}gen 6B, SE-41296 Gothenburg, Sweden}
\author{Harshith Bachimanchi}
\affiliation{Department of Physics, University of Gothenburg, Origov{\"a}gen 6B, SE-41296 Gothenburg, Sweden}
\author{Sergio No\'{e}}
\affiliation{Facultat de Ci\`encies i Tecnologia, Universitat de Vic -- Universitat Central de Catalunya (UVic-UCC), C. de la Laura,13, 08500 Vic, Spain}
\author{Daniel Midtvedt}
\affiliation{Department of Physics, University of Gothenburg, Origov{\"a}gen 6B, SE-41296 Gothenburg, Sweden}
\author{Giovanni Volpe}
\email{giovanni.volpe@physics.gu.se}
\affiliation{Department of Physics, University of Gothenburg, Origov{\"a}gen 6B, SE-41296 Gothenburg, Sweden}
\author{Carlo Manzo}
\email{carlo.manzo@uvic.cat}
\affiliation{Facultat de Ci\`encies i Tecnologia, Universitat de Vic -- Universitat Central de Catalunya (UVic-UCC), C. de la Laura,13, 08500 Vic, Spain}
\date{\today}

\begin{abstract} 
The characterization of dynamical processes in living systems provides important clues for their mechanistic interpretation and link to biological functions. Thanks to recent advances in microscopy techniques, it is now possible to routinely record the motion of cells, organelles, and individual molecules at multiple spatiotemporal scales in physiological conditions. However, the automated analysis of dynamics occurring in crowded and complex environments still lags behind the acquisition of microscopic image sequences. Here, we present a framework based on geometric deep learning that achieves the accurate estimation of dynamical properties in various biologically-relevant scenarios. This deep-learning approach relies on a graph neural network enhanced by attention-based components. By processing object features with geometric priors, the network is capable of performing multiple tasks, from linking coordinates into trajectories to inferring local and global dynamic properties. We demonstrate the flexibility and reliability of this approach by applying it to real and simulated data corresponding to a broad range of biological experiments. 
\end{abstract}

\keywords{deep learning; motion; tracking; graph neural network; migration; diffusion}

\maketitle

\section*{Introduction}

The biological functions of living systems rely on interactions that dynamically change in response to endogenous and exogenous stimuli. 
Studying the motion of the individual components of these systems sets the basis for mechanistic insights to understand health and disease~\cite{bruckner2021learning}. 
Over the last 20 years, microscopy has advanced to the point where it can monitor dynamic processes at multiple scales with unprecedented spatiotemporal resolution. Time-lapse microscopy experiments have unveiled the strategies that unicellular organisms employ to search for food or to avoid adverse conditions, and have helped to understand tissue growth and repair, cancer metastasis, quorum sensing, the emergence of multicellularity, and immune responses in multicellular organisms~\cite{ladoux2017mechanobiology,ramos2021environment}. 
Fluorescence microscopy has monitored biological motion down to the nanoscale, detailing the diffusion of individual organelles and molecules within the cellular environment and disclosing their role, e.g., in the fundamental processes of signaling and function regulation~\cite{manzo2015review, shen2017single}. 
Tethered-particle microscopy as well as optical and magnetic tweezers have used the motion of micron-sized beads as a proxy to infer changes in the kinetics of proteins and nucleic acids at the single-molecule level~\cite{gieseler2021optical}. 

\begin{figure*}[hbt!]
    \centering
    \includegraphics[width=1\textwidth]{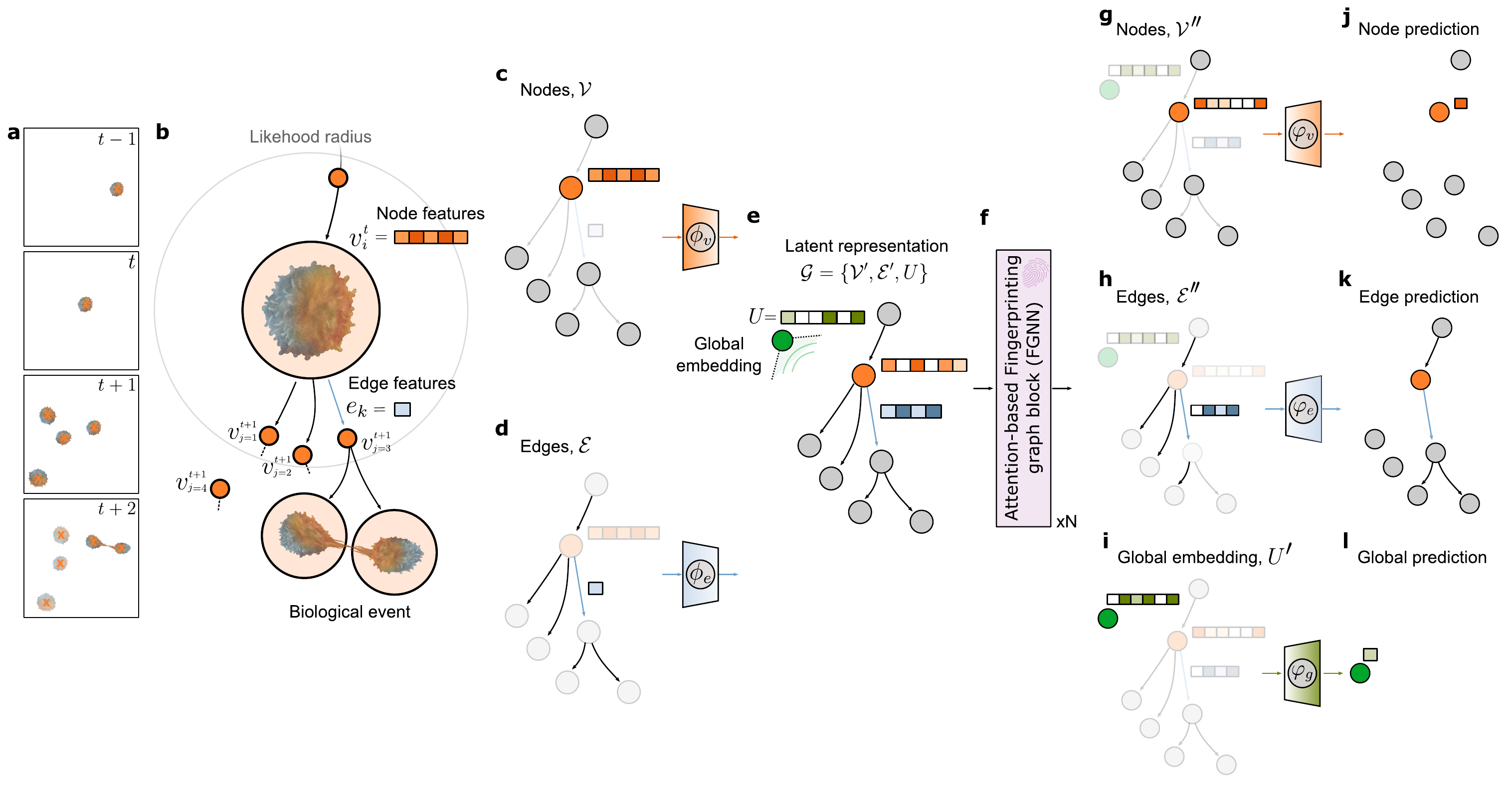}
    \caption{\textbf{Spatiotemporal characterization of trajectories using MAGIK}. 
    \textbf{a}, Sequence of images illustrating the evolution of a group of cells over time, corresponding to frame numbers $t-1$, $t$, $t+1$, $t+2$. The orange crosses indicate a detection. 
    \textbf{b}, The movement of the cells and their interactions are geometrically modeled using a directed graph, where the nodes ($\mathcal{V}$) represent the detections and the edges ($\mathcal{E}$) connect spatiotemporally-close detections. Each node contains features (orange squares) such as the cell's centroid and some relevant descriptors (e.g., the cell's morphological and intensity attributes). The edges contain features (blue squares) too, in this case encoding the Euclidean distance between the centroids of the cells. In this example, the node of interest, labeled with the subindex $i$, is connected to neighboring nodes in the future, labeled with the subindex $j$ within a distance-based likelihood radius (the edge between nodes $v_i^{t}$ and $v_{j=4}^{t + 1}$ is dumped). Meaningful biological events (e.g., cell divisions) are naturally encoded in the graph. 
    \textbf{c, d,} The input node and edge features are mapped to a higher-level feature representation using learnable encoding functions implemented by the neural networks $\phi_v$ and $\phi_e$, respectively.
    \textbf{e} Importantly, we also append an extra learnable token $U$ to the graph latent representation $\mathcal{G} = \{\mathcal{V}', \mathcal{E}', U\}$, whose function is to provide global insights about the dynamics of the cells. \textbf{f}, MAGIK relies on attention-based fingerprinting graph blocks (FGNN) to process $\mathcal{G}$ and provide an updated representation for nodes ($\mathcal{V}^{''}$, \textbf{g}), edges ($\mathcal{E}^{''}$, \textbf{h}), and global information ($U^{'}$, \textbf{i}) (for further details regarding the FGNN architecture, refer to Methods, ``Description of MAGIK'', and Suppl.~Fig.~\ref{fig:s1}). 
    Finally, $\mathcal{V}^{''}$, $\mathcal{E}^{''}$, and $U^{'}$ are decoded by applying learnable functions implemented by the neural networks $\varphi _v$, $\varphi _e$, and $\varphi _u$, respectively, to obtain the sought-after node (\textbf{j}), edge (\textbf{k}), and global information (\textbf{l}).}
    \label{fig:1}
\end{figure*}

The momentous improvement of microscopy acquisition techniques has led to a substantial effort to develop and improve algorithms to automatically extract quantitative information from these experiments~\cite{chenouard2014objective, ulman2017objective}. The standard analysis pipeline of tracking-by-detection methods entails the following steps~\cite{manzo2015review, ulman2017objective, tinevez2017trackmate}: 
1. Objects of interest are detected in movie frames (segmentation). 
2. Object positions and other state parameters are estimated (localization).
3. Detected positions at different times are connected into trajectories (linking). 
4. Reconstructed trajectories are finally analyzed to quantify dynamical parameters (estimation).
The first 3 steps are often presented together and referred to as tracking. In contrast, likely due to the broad variety of different parameters it might be required to evaluate, the estimation step is usually considered separately.
For biological experiments, this analysis is made more difficult by various factors, such as imaging noise, high object density, fusion or splitting events, random and heterogeneous motion, and shape-changing objects. 
Errors at each step propagate along the pipeline and ultimately impact the extraction of dynamic information. 

Several algorithmic solutions have been proposed to tackle limitations of tracking algorithms and their performance has been compared in open challenges~\cite{chenouard2014objective, ulman2017objective}. However, most of these methods are specific to a given experiment or dynamic model, and often require manual tuning of parameters. 
The current deep-learning revolution has fostered the development of various methods for both tracking~\cite{helgadottir2019digital,berg2019ilastik,midtvedt2021quantitative,ershov2021bringing} and estimation~\cite{munoz-gil2021objective}. However, deep-learning-powered approaches have so far been bound to follow the standard analysis pipeline, providing data-driven versions of conventional approaches without taking full advantage of their possibilities~\cite{ershov2021bringing}. 

Geometric deep learning provides compelling approaches to tackle tracking and estimation from a different perspective.
It generalizes neural networks to problems that can be described by mathematical objects such as graphs that encode information about the structure of the input~\cite{bronstein2017geometric}. Deep learning methods based on graphs are typically referred to as graph neural networks (GNNs)~\cite{battaglia2018relational} and have been successfully applied, e.g., to molecular property prediction~\cite{liu2019chemi}, drug discovery~\cite{stokes2020deep}, and computer-assisted retrosynthesis~\cite{somnath2021learning}. Besides being ubiquitously used in science to represent complex systems~\cite{strogatz2001exploring}, graphs provide a natural and intuitive way to represent the information contained in tracking experiments~\cite{loffler2021graph, verdier2021learning}. 

Here, we describe a framework for Motion Analysis through GNN Inductive Knowledge (MAGIK), which provides the accurate estimation of dynamical properties from time-lapse microscopy.
MAGIK models the system's motion and interactions through a graph representation. This graph is processed through an interpretable and adaptive attention-based GNN that estimates the associations among the objects and provides insights into the intrinsic dynamics of the systems. 
We demonstrate the flexibility and reliability of MAGIK by quantifying its performance on real and simulated data corresponding to a broad range of biological experiments. 
First, we benchmark it on its most natural application, i.e., trajectory linking, in a variety of challenging experimental scenarios, including high-density experiments, merge/split events, and shape-changing objects, where MAGIK features gap-closing capabilities, gracefully handles segmentation errors, and even overrides imperfect annotation of training datasets.
Then, going beyond optimal trajectory linking, we show that MAGIK can further estimate local and global dynamical properties, such as diffusion coefficients, diffusion modes, and anomalous diffusion exponents, even in highly heterogeneous scenarios at the ensemble and single-object levels. 

\section*{Results}

\subsection*{MAGIK represents spatiotemporal relations in a graph.}

MAGIK provides a GNN framework to estimate the dynamical properties of moving objects from time-lapse experiments, which are relevant in different biological scenarios. MAGIK models the objects' motion and physical interactions using a graph representation. The details of the algorithm are given in \hyperlink{sec:methods}{Methods} (``Description of MAGIK") and Fig.~\ref{fig:s1}. In this section, we provide a high-level description of the architecture, as represented in Fig.~\ref{fig:1}.  

Graphs can define arbitrary relational structures between nodes connecting them pairwise through edges. 
When training a GNN, the graph architecture guides the learning process about the objects and their relations by introducing a relational inductive bias~\cite{battaglia2018relational}.
In MAGIK, each node describes an object detection at a specific time, the edges connect spatiotemporally close objects, and a set of global attributes encodes system-level properties. As an example, for subsequent frames of a cell migration experiment, each detected object (orange crosses in Fig.~\ref{fig:1}a) is associated with a node with a vector of node features (Fig.~\ref{fig:1}b). Directed edges with relational features connect each node to objects detected in the future in its proximity (Fig.~\ref{fig:1}b). There are no intrinsic restrictions on the type or number of descriptors (e.g., location and morphological features, image-based quantities, biological events, interaction strength, distance, direction) that can be encoded in the graph feature representation. The basic graph relational structure is established through a set of rules that link nodes pairwise based on distance metrics between features. Node and edge features are encoded through learnable functions implemented by neural networks (Figs.~\ref{fig:1}c, d). An extra learnable token is added to aggregate global attributes from the whole graph~\cite{dosovitskiy2020image} (Fig.~\ref{fig:1}e).
\begin{figure*}[hbt!] 
    \centering
    \includegraphics[width=0.46\textwidth]{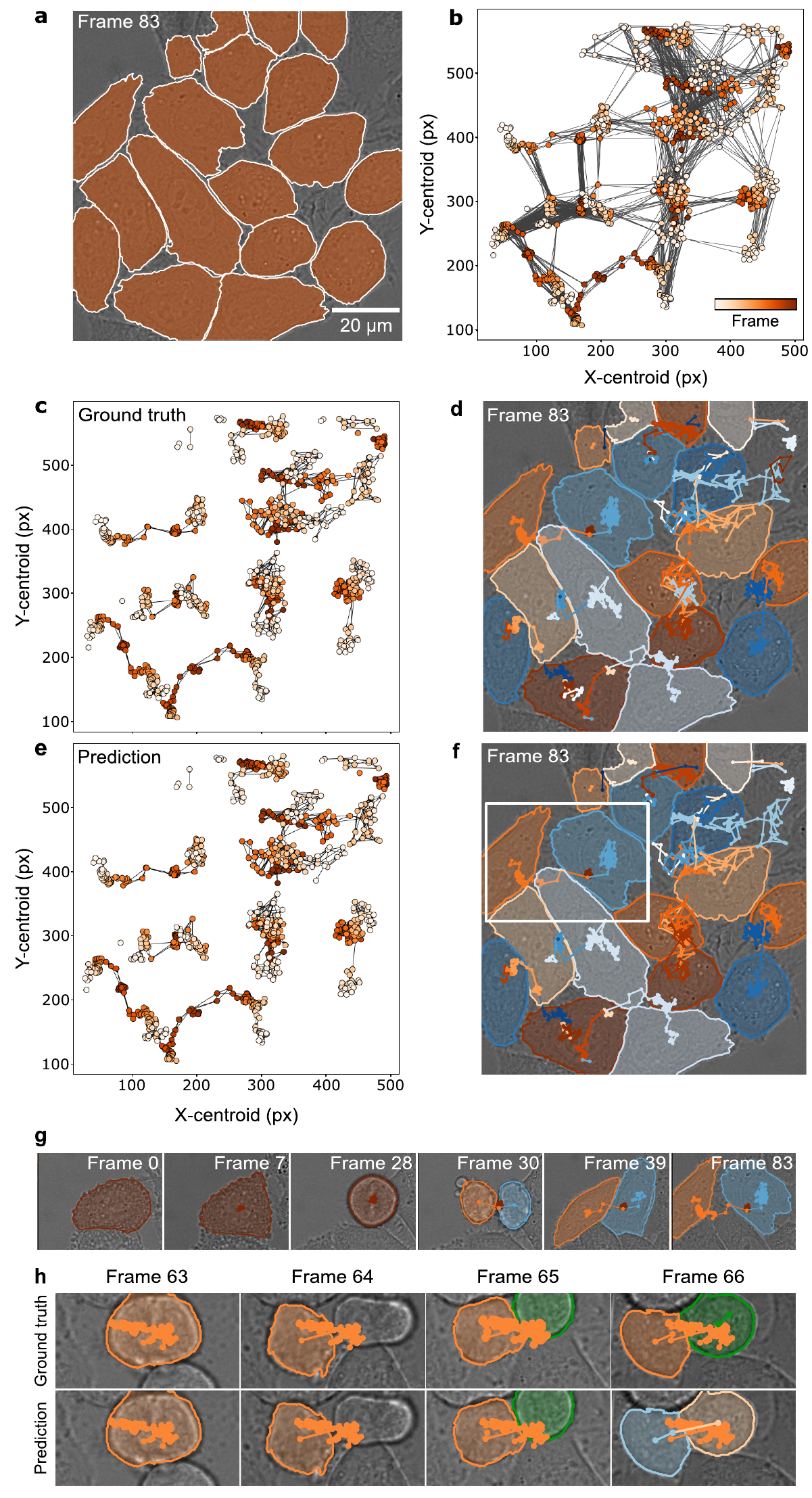}
    \caption{
    \textbf{Trajectory linking using MAGIK}. 
    \textbf{a}, Representative frame of the HeLa cells in the DIC-C2DH-HELA validation video of the $6^{\rm th}$ Cell Tracking Challenge ~\cite{ulman2017objective}. Segmentations (colored regions) are used to extract relevant information from each cell along the sequence of images to build \textbf{b}, the input graph structure including a redundant number of edges with respect to the actual associations between objects.
    \textbf{c}, Ground-truth graph, and \textbf{d}, ground-truth cell trajectories. 
    \textbf{e}, The predicted graph agrees well with the expected solution, achieving a $\textup{F}_1$-score equal to $99.4\%$. 
    \textbf{f}, The predicted trajectories reach $\operatorname{TRA}=99.2\%$ compared to the ground-truth. 
    Cell divisions are detected correctly, and the network performs well also in edge regions where cells are only partially observed and move out of the field of view.
    \textbf{g}, Zoomed-in view of the inset in \textbf{f} showing the heterogeneity in cell shape and dynamics. A cell changes morphology during migration (frame 0-28) and divides into 2 daughter cells (frame 30) that spread and migrate apart (frames 30-83).
    \textbf{h}, MAGIK exhibit emergent gap-closing capabilities. 
    MAGIK is able to identify the white and blue cells at frame 66 as the product of a cell division despite the missing segmentation of one of the daughter cells (frame 64) that led to the misclassification of the green cell (frames 65-66) in the annotation of ground-truth trajectory.
    A video visualizing the tracked cells can be found in the supplementary material (Suppl. Video~\ref{vid:s1})
    }
    \label{fig:2}
\end{figure*}

The graph is processed through a sequence of attention-based fingerprinting graph neural networks (FGNN, see also ``Description of MAGIK'' and Suppl.~Fig.~\ref{fig:s1}) that propagate information through the graph via message-passing steps (Figs.~\ref{fig:1}f-i).
The relational inductive knowledge implemented in the graph structure sketches a network of redundant object associations. The objective of the FGNN is to modulate the association strength to identify the edges majorly influencing the dynamic properties of each object.
For this, the FGNN implements two attention mechanisms that combine information from multiple objects while considering the presence of heterogeneity at the local and global levels. The first mechanism intervenes when aggregating edge features to a node (Eq.~\ref{eq:h_i}). The contribution of each edge has a weight that depends on the distance between the connected nodes through a function with learnable parameters (Eq.~\ref{eq:w_ij}), thus defining a learnable local receptive field that allows the network to adapt to heterogeneous dynamics and to robustly account for relevant relations between the nodes. The second is a gated self-attention mechanism~\cite{jumper2021highly} that sets in when updating the latent representation of nodes (Eq.~\ref{eq:attn}). The node update operation involves also information stemming beyond each node's topological neighborhood, thus effectively expanding the receptive field to objects that, although not physically connected, can offer relevant information about the overall dynamics. The use of gated self-attention offers a feature-wise discriminatory power to the node update operation since it weights individual features of the attention node embedding with respect to their importance to the overall graph structure. Through this mechanism, MAGIK identifies only the meaningful features of each node. Unreliable or incomplete features (e.g., the morphology of objects at the edge of the image or partially outside the field of view) are thus prevented from adding noise to the correct prediction of the network. 
The FGNN further updates the extra token for global attributes using information from all the nodes; thus, this extra token serves as an antenna to provide system-level insights. 

The output of the FGNN is decoded by the last block of the GNN into an output graph, whose nodes, edges, and global attributes can be used to solve specific problems (Figs.~\ref{fig:1}j-l).  The flexibility of the graph representation and the possibility to use various types and numbers of input features make MAGIK suitable for determining multiple parameters associated with various experimental scenarios where there are objects of interest moving in space and time. 

In the following sections, we exemplify the application of MAGIK to: (i) the analysis of experiments of cell migration to determine trajectories in the presence of proliferation; (ii) fluorescence imaging of single molecules to determine parameters of heterogeneous and anomalous diffusion; and (iii) holographic imaging of microorganisms to classify their diffusion mode. In all cases, MAGIK is trained for $\approx 100$ epochs with small datasets (a single video of $\approx 100$ frames for linking; at most $\approx 1000$ videos of $\approx 50$ frames each for the other cases) thanks to the use of an ad hoc augmentation procedure that combines feature corruption and node dropping, thus enabling transfer learning to more challenging conditions with no loss of performance. The training is typically completed in minutes on a GPU-enhanced computer (see \hyperlink{sec:methods}{Methods}, ``MAGIK training").

\subsection*{MAGIK accurately links trajectories.}

We first benchmark MAGIK performance on a classical trajectory linking task, consisting of establishing temporal associations between identified objects. 
For object linking, the graph structure includes a redundant number of edges with respect to the actual associations between objects. The aim of MAGIK is to prune the wrong edges while retaining the true connections by using all the available spatiotemporal information. 
We thus model this task as an edge-classification problem with a binary label (linked/unlinked). From the predicted edge features, trajectories are built through a postprocessing algorithm that eliminates spurious connections (\hyperlink{sec:methods}{Methods}, ``Postprocessing algorithm for trajectory linking").
 
To test MAGIK, we use the silver-standard segmentation datasets provided for the training of the sixth edition of the Cell Tracking Challenge~\cite{ulman2017objective} (which has been created by combining results of several automatic analysis methods following a majority-voting scheme). 
A representative segmentation of the dataset DIC-C2DH-HELA, corresponding to HeLa cells on a flat glass imaged through differential interference contrast, is shown in Fig.~\ref{fig:2}a. From the segmentation, we calculate the mean pixel intensity, area, perimeter, eccentricity, and solidity of the segmented objects, which we use as input node features. The Euclidean distance between neighboring objects is used as the sole edge feature. To limit memory usage, we generate graphs by drawing edges only between objects within a limited spatial and temporal reach (Fig.~\ref{fig:2}b).  

The DIC-C2DH-HELA dataset presents several challenges, namely, high packing density, low signal-to-noise ratio, and a highly heterogeneous intracellular signal due to DIC-highlighted internal structures and organelles. The heterogeneity further extends to cell shape and dynamics over time as a consequence of migration and proliferation (Figs.~\ref{fig:2}g,h). Examples of ground-truth and predicted graphs are shown in Figs.~\ref{fig:2}c,e showing a good agreement, as confirmed by a $\textup{F}_1$-score of $99.4\%$ in edge prediction. For the evaluation of performance at the trajectory level (Figs.~\ref{fig:2}d,f), we calculated the tracking accuracy measure ($\operatorname{TRA}$), a normalized weighted distance between the tracking prediction and the reference tracking ground truth~\cite{matula2015cell} (\hyperlink{sec:methods}{Methods}, ``Quantification of cell tracking results"). When evaluated with respect to trajectories, MAGIK reached a $\operatorname{TRA}=99.2\%$ showing a great capability of correctly following objects despite imperfect segmentation, shape changes, and cell divisions (Fig.~\ref{fig:2}g, Suppl. Video~\ref{vid:s1}). 

An interesting emergent capability of the method is highlighted in Fig.~\ref{fig:2}h. The video microscopy at frame 63 captures a cell (orange shadow) dividing into two daughter cells (frames 64-66). The ground-truth segmentation at frame 64 misses one of the daughter cells, preventing the identification of the division event at this frame. This kind of error is not uncommon because, for actual experiments, although most of the annotations can be considered as true positives they are a subset of the unknown ground truth. 
When the second daughter cell is identified (green shadow at frame 65), both MAGIK and the annotations used as ground-truth associate it to a new trajectory. However, at frame 66, MAGIK is able to identify the white and blue cells as the product of the cell division of the orange cell at frame 65, highlighting a general learning ability of the network based on the propagation of topological and morphological information over time.
Although the detection of these events is not reflected in the computation of tracking metrics, their identification is relevant for the biological interpretation of the experiment (e.g., calculating the cell division rate)~\cite{ulman2017objective}.
\begin{figure*}[hbt!]
    \centering
    \includegraphics[width=0.8\textwidth]{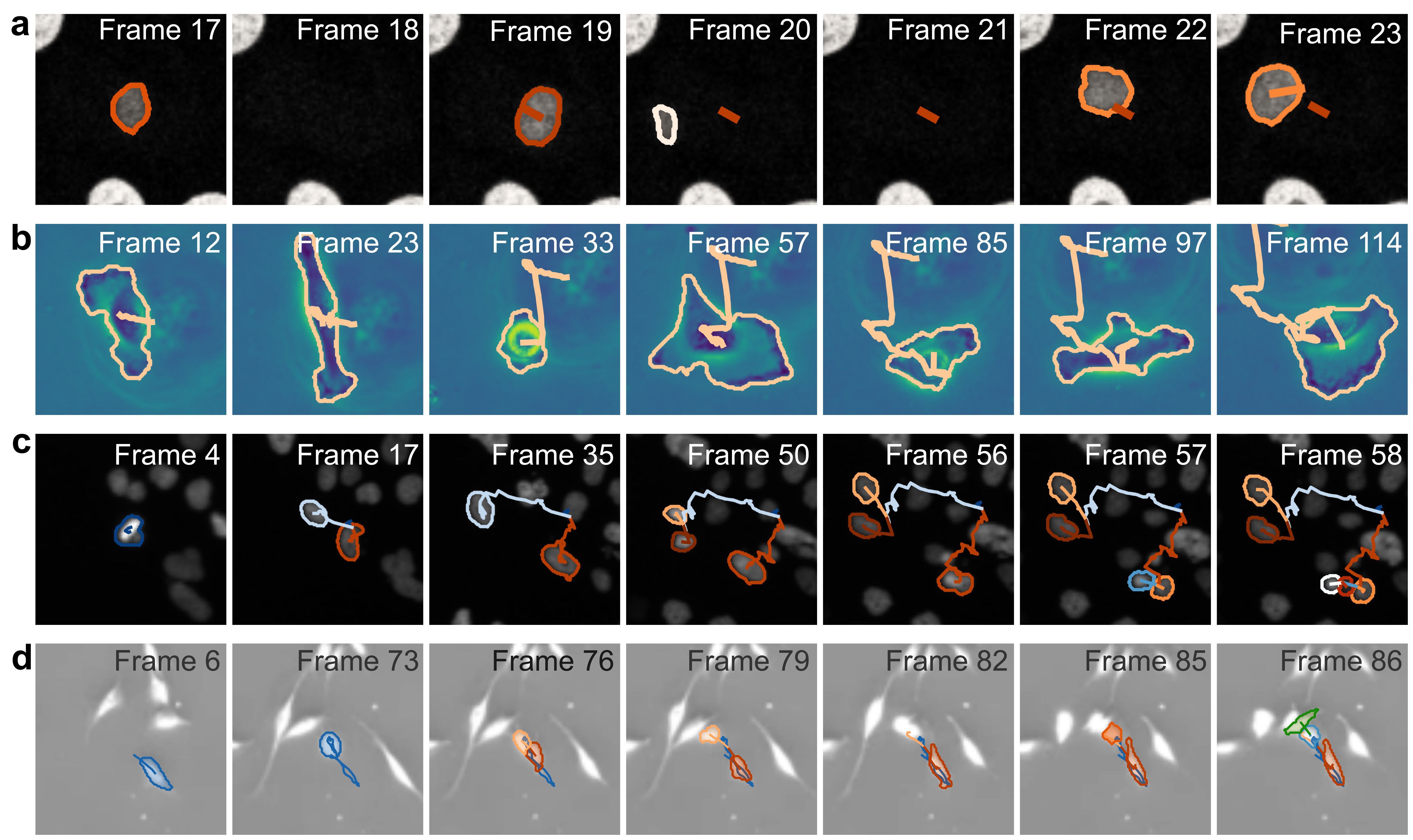}
     \caption{
     \textbf{MAGIK reliably links trajectories in various experimental scenarios}.
     \textbf{a}, Confocal microscopy of GFP-GOWT1 mouse stem cells. MAGIK achieves $\textup{F}_1$-score = $99.8\%$ and $\operatorname{TRA}=99.2\%$ despite the fact that the cells frequently leave the field of observation.
     \textbf{b}, Phase-contrast imaging of glioblastoma-astrocytoma U373 cells on a polyacrylamide substrate. MAGIK reaches $\textup{F}_1$-score = $99.8\%$ and $\operatorname{TRA}=100\%$ even though the cells greatly change shape over time. 
     \textbf{c}, Epifluorescence imaging of HeLa cells stably expressing H2b-GFP. MAGIK achieves $\textup{F}_1$-score = $98.8\%$ and $\operatorname{TRA}=98.4\%$ despite the dense sample and frequent mitosis and collisions.
     \textbf{d}, Phase-contrast imaging of pancreatic stem cells on a polystyrene substrate.  MAGIK obtains $\textup{F}_1$-score = $99.3\%$ and $\operatorname{TRA}=98.5\%$ despite high cell density, elongated shapes, pronounced cell displacements, and a significant number of division events.
     Interrupted trajectories correspond to cases where cells left the field of view or missed segmentation in the image sequence.
     All videos belong to the dataset of the $6^{\rm th}$ Cell Tracking Challenge ~\cite{ulman2017objective}
     Results can be observed in greater detail in Supplementary Videos~\ref{vid:s2}-\ref{vid:s5}.
     }
    \label{fig:3}
\end{figure*}

We applied MAGIK to several other datasets of the $6^{\rm th}$ Cell Tracking Challenge obtaining outstanding results for different microscopy techniques and cell types. Representative video frames with segmentation are shown in Fig.~\ref{fig:3} for confocal microscopy imaging of GFP-GOWT1 mouse stem cells (Fig.~\ref{fig:3}a, $\textup{F}_1$-score = $99.8\%$, $\operatorname{TRA}=99.2\%$), phase-contrast imaging of glioblastoma-astrocytoma U373 cells on a polyacrylamide substrate (Fig.~\ref{fig:3}b, $\textup{F}_1$-score = $99.8\%$, $\operatorname{TRA}=100\%$), epifluorescence imaging of HeLa cells stably expressing H2b-GFP (Fig.~\ref{fig:3}c, $\textup{F}_1$-score = $98.8\%$, $\operatorname{TRA}=98.4\%$), and phase-contrast imaging of pancreatic stem cells on a polystyrene substrate (Fig.~\ref{fig:3}d, $\textup{F}_1$-score = $99.3\%$, $\operatorname{TRA}=98.5\%$) (see Suppl. Videos~\ref{vid:s2}-\ref{vid:s5} for full movies). Even though a strict objective comparison of MAGIK linking capability with other methods is limited by the fact that different algorithms rely on different segmentations (whose errors influence linking and thus indirectly affect the value of the $\operatorname{TRA}$ metric), MAGIK obtained $\operatorname{TRA}$ values that are competitive, if not superior, to the best-in-class methods of the $6^{\rm th}$ Cell Tracking Challenge.

\subsection*{MAGIK quantifies motion parameters without trajectory linking.}

In most applications, the ultimate objective of tracking is the characterization of the dynamics of the systems under investigation to gain insights into their underlying biological mechanisms.
In this process, trajectory linking is often just an intermediate step necessary to get meaningful information from the data, but not the end goal itself. 
For example, in single-molecule fluorescence microscopy, trajectory analysis is often performed to quantify dynamic parameters such as diffusion coefficients, to determine the extent of mixed diffusive behaviors (e.g., slow/fast, mobile/confined), and to classify the diffusion mode~\cite{manzo2015review, shen2017single,munoz-gil2021objective}.

Differently from most other estimation techniques, MAGIK can characterize essentially any dynamic aspect of the system under investigation without requiring the actual linking, thanks to its capability of accounting for the whole spatiotemporal complexity contained in the associations between objects at multiple scales.  
Such linking-free analysis produces a two-fold advantage. First, it bypasses the error-prone linking step, thus inherently preventing linking errors from propagating to the quantification of the ultimately relevant parameters. 
Second, it enables the analysis of experiments for which linking cannot be performed due to, e.g., a high object density or low signal-to-noise ratio.

\begin{figure*}[hbt!]
    \centering
    \includegraphics[width=1\textwidth]{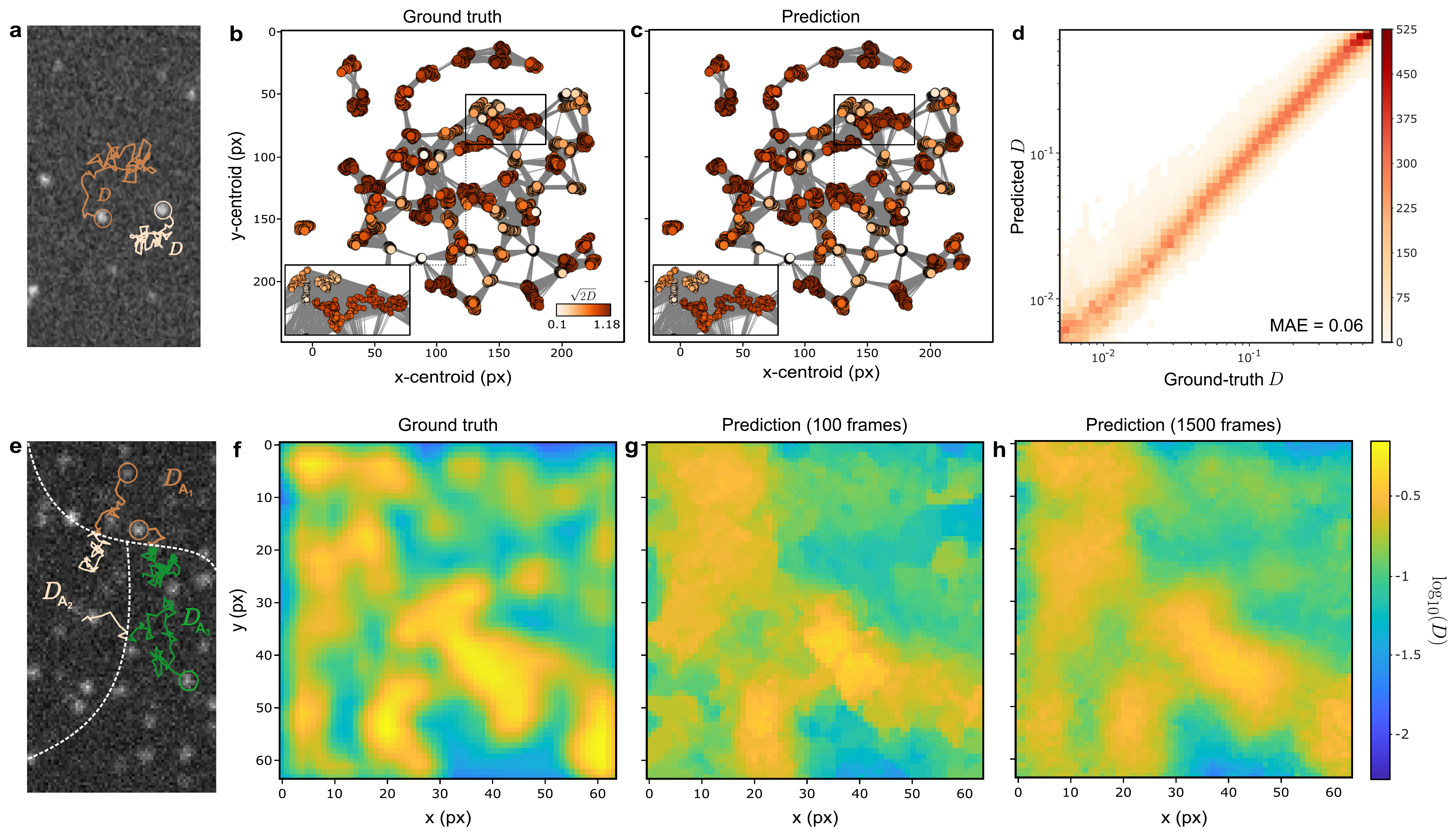}
    \caption{
    \textbf{MAGIK determines local diffusion properties.}
    \textbf{a}, Simulated single-object tracking experiment where fluorescence microscopy is used to follow the motion of single molecules performing Brownian motion with diffusivity $D$ varying from particle to particle. 
    \textbf{b, c}, Ground-truth and predicted graphs. The edges depict the network of associations used to infer dynamic properties without direct linking. The nodes are color-coded according to the value of the target feature, i.e. the displacement scaling factor $\sqrt{2D}$ measured in pixels per frame (color bar in \textbf{b}). 
    \textbf{d} Probability distribution of the predicted vs. ground-truth diffusion coefficient $D$, showing a good agreement.
    \textbf{e}, Simulated single-object tracking experiment where fluorescence microscopy is used to follow the motion of single molecules performing Brownian motion with diffusivity $D$ randomly varying in space.
    \textbf{f, g, h}, Ground-truth and predicted diffusion maps. Ground-truth spatial diffusivity pattern (\textbf{f}) and prediction obtained by MAGIK using a 100- (\textbf{g}) and 1500-frame-long (\textbf{h}) movies with $\approx0.02$ localizations per px$^{2}$ per frame. The analysis is performed by breaking down the sequence in 2 and 30 videos of 50 frames each, respectively. 
    Predicted maps are obtained by interpolating the values of diffusivity obtained for the nodes over the $64\,{\rm px}\times 64\,{\rm px}$ grid through a triangulation-based nearest-neighbor algorithm. 
    }
    \label{fig:4}
\end{figure*}

To highlight its capabilities and quantitatively assess its performance, in Fig.~\ref{fig:4}, we apply MAGIK to analyze simulated data reproducing the diffusion of fluorescently-labeled single molecules like, e.g., lipids or receptors in the plasma membrane of living cells. 
We first consider the task of determining the diffusion coefficient from a heterogeneous ensemble of diffusing objects (Fig.~\ref{fig:4}a). 
We feed the network the centroid coordinates and intensity of the localized fluorescence spots as node features and the Euclidean distance between neighboring centroids as the edge feature. We define the problem as a node regression where the target feature is the displacement scaling factor $\sqrt{2D}$, with $D$ being the diffusion coefficient of the molecule associated with each node. Graphs are built by connecting localized objects with neighbors in space and time (Fig.~\ref{fig:4}b).  Ground-truth and predicted graphs are shown in Fig.~\ref{fig:4}b and Fig.~\ref{fig:4}c, respectively.  All the edges of the graph structure are drawn, representing the network of associations used to infer dynamic properties without direct linking. Nodes are color-coded according to the value of the displacement scaling factor $\sqrt{2D}$.  Their visual comparison suggests excellent agreement, further confirmed by the quantification in Fig.~\ref{fig:4}d. 
The same approach can also be extended to estimate other parameters. In Suppl. Fig.~\ref{fig:s2}a-d, we show the results of its application to the inference of the scaling exponent for objects undergoing anomalous diffusion, achieving  similarly good results.
Fluorescence microscopy experiments for single-object tracking must ensure that the number of visualized molecules is low enough to unambiguously link the trajectories, thus are often performed at low labeling density ~\cite{manzo2015review}. However, these conditions are not optimal to probe the interactions between particles and make difficult the inference of spatial patterns of diffusion~\cite{jaqaman2008robust}. Enabling the quantification of diffusion properties without linking offers the possibility to process high-density videos to determine the underlying topology and spatial heterogeneity. 
As an example, we used MAGIK to resolve a spatially-modulated landscape with diffusion continuously varying over more than 2 orders of magnitude from the localizations of diffusing particles (Fig.~\ref{fig:4}e-h), treating the problem as a node feature regression, as above. At a number density of $\approx0.02\,{\rm px^{-2}}$, about one order of magnitude higher  than the limit for reliable tracking~\cite{chenouard2014objective},
MAGIK is capable of correctly retrieving the spatial map of $D$ (Fig.~\ref{fig:4}f). Remarkably, most spatial features can be already resolved with a 100-frames long movie (Fig.~\ref{fig:4}g). 
The spatial resolution of the predicted map can be further improved using longer videos (1500 frames, Fig.~\ref{fig:4}h), with the typical duration of single-molecule fluorescence microscopy experiments for measuring diffusion \cite{manzo2015review}.

\subsection*{MAGIK quantifies global dynamic properties.}

\begin{figure*}[hbt!]
    \centering
    \includegraphics[width=1\textwidth]{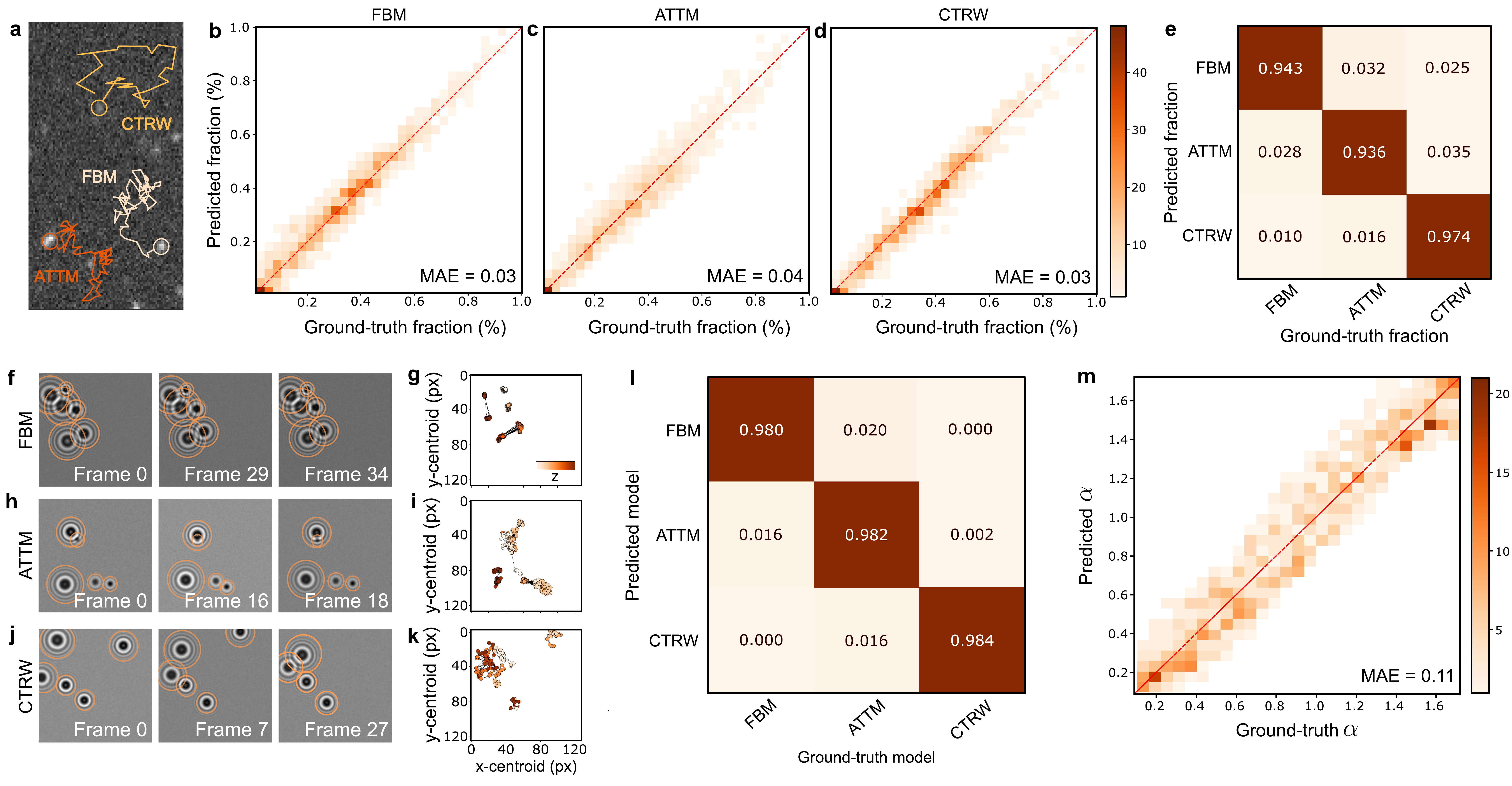}
    \caption{
    \textbf{MAGIK estimates local and global dynamic properties at the ensemble and single-object levels.} 
    \textbf{a}, Simulated single-object tracking experiment where objects with different underlying diffusion models coexist (i.e., fractional Brownian motion (FBM), annealed transient time motion (ATTM), and  continuous-time random walk (CTRW) with anomalous diffusion exponents $\alpha = 1$).
    \textbf{b-d} Probability distribution of predicted vs. ground-truth model fraction for FBM, ATTM, and CTRW, respectively.
    \textbf{d} Confusion matrix demonstrating how the network classifies the underlying diffusion model exhibited by objects in 1199 validation videos. 
    The diagonal represents the percentage of correctly classified graph representations, constituting most cases. The off-diagonal cells represent incorrectly classified examples.
    Column-based normalization is applied, such that the sum along the columns adds up to 1, with minor deviations due to rounding.
    \textbf{f-k}, Representative frames of simulated holographic video and corresponding graph representation for the whole image sequence, where objects follow \textbf{f-g} FBM, \textbf{h-i}, ATTM, and \textbf{j-k} CTRW, with $\alpha = 1$. In the graphs, the edges depict the association network used to infer dynamic properties without trajectory linking. 
    \textbf{l}, Confusion matrix showing how the network classifies the underlying diffusion model presented in 1496 validation videos. Column-based normalization is applied. 
    \textbf{m}, MAGIK predicts the anomalous diffusion exponent governing the motion of ensembles of objects performing FBM in 1097 holographic videos. The probability distribution of the predicted  vs. ground-truth anomalous diffusion exponent ($\alpha$) exhibits a good performance throughout the evaluated range.
    }
    \label{fig:5}
\end{figure*}

We further applied MAGIK to extract ensemble information through the inference of global attributes skipping direct trajectory linking in two biologically-relevant experimental scenarios. As a first example, we considered fluorescence microscopy experiments in which objects in the same video undergo diffusion according to different microscopic models (namely, fractional Brownian motion (FBM), annealed transient time motion (ATTM), and continuous-time random walk (CTRW), Figs.~\ref{fig:5}a-e). Although these diffusion models can give rise to anomalous diffusion, in this example they are parametrized so to have the same scaling of the mean-squared displacement of Brownian motion ($\alpha = 1$)~\cite{munoz-gil2021objective}. 
Graphs are built as described above using centroid coordinates and intensity of the localized fluorescence spots as node features and the Euclidean distance between neighboring centroids as the edge feature. MAGIK estimates the relative fraction of objects in each category, varying from experiment to experiment, as a regression problem on the global attribute. Results obtained over a large set of experiments are summarized in Fig.~\ref{fig:5}a-e, showing an outstanding accuracy in predicting the correct fractions, even when the number of objects performing the same class of motion in the experiment is very low.
In Suppl. Fig.~\ref{fig:s2}e-h, we further demonstrate that the same approach can also estimate the fraction of object  moving according to different diffusion modes (subdiffusion with $\alpha<1$, normal diffusion with $\alpha=1$, and superdiffusion with $\alpha>1$).

The second example refers to simulations of holographic imaging of microorganisms diffusing in a liquid environment, such as plankton (Fig.~\ref{fig:5}f-k). In this case, we model diffusion as either FBM (Fig.~\ref{fig:5}f-g), ATTM (Fig.~\ref{fig:5}h-i), or CTRW (Fig.~\ref{fig:5}j-k) with $\alpha=1$. Objects in the same experiments move according to the same physical model but with random diffusivity. Centroid 3D coordinates, mean intensity, area, and refractive index of the objects are used as node features in a classification problem to determine the common diffusion model of the objects in the same video, encoded as a global attribute.  As shown in Fig.~\ref{fig:5}l, MAGIK correctly classifies the generative dynamics even with largely overlapping objects. We find this result quite remarkable (equally so as that illustrated in Suppl. Fig.~\ref{fig:4}e) since, for $\alpha=1$, all models converge to Brownian motion and feature large similarities in their statistical properties, making their classification rather challenging even when linked trajectories are available~\cite{munoz-gil2021objective}. 

Last, we explore MAGIK's performance for quantifying anomalous diffusion through the estimation of the exponent $\alpha$~\cite{munoz-gil2021objective} from a sequence of holographic images reproducing the motion of microorganisms. All the objects in the same movie undergo FBM with random diffusivity and the same exponent $\alpha$, varying from sequence to sequence (Fig.~\ref{fig:5}m). Also in this case, MAGIK provides remarkable results (MAE$=0.11$) from short movies ($\approx 50$ frames) containing only a few objects.
\section*{Discussion}

MAGIK is a general and versatile approach for the characterization of dynamic properties from time-lapse microscopy that exploits geometric deep learning capability to capture the full spatiotemporal complexity of biological experiments. MAGIK strongly relies on an attention-based GNN that can extract dynamic parameters from image-based features by assuming relational constraints between objects. 
The use of these relational properties at multiple scales makes MAGIK robust with respect to missed detections, object appearance/disappearance, merge/split events, and high object densities.
In this article, we have shown several applications that highlight its most important features in different biological contexts.

For trajectory linking, currently used methods employ Kalman filter~\cite{godinez2011tracking}, multiframe and/or multitrack optimization based on greedy algorithms that approximate the multiple-hypothesis tracking (MHT) solution~\cite{jaqaman2008robust, chenouard2009multiple, coraluppi2011multi}, or combinatorial optimization~\cite{sbalzarini2005feature}. Most of these approaches offer their best performance when knowledge of the motion is explicitly used~\cite{chenouard2014objective}. MAGIK provides the first efficient data-driven alternative to these approaches and can be used to analyze any kind of motion and interaction pattern. Training can be performed on a minimal amount of annotated data or simulations. The inductive bias encoded in the graph structure offers an inherent gap-closing capability and reduces the combinatorial space of potential solutions. Flexible hyperparameters allow the tuning of the maximum distance and time lag for node connection to avoid missed links even for highly heterogeneous motion.

MAGIK provides a key enabling technology to estimate dynamic parameters from segmentation/localization in a complete linking-free fashion, whereas other methods require some level of knowledge about the linking between objects~\cite{el2015inferencemap, xiang2020single}. As such, it provides a powerful solution for those experiments where trajectory linking cannot be reliably performed, e.g., as a consequence of high object density or probe blinking. By the inference of node properties, we demonstrate capability for resolving spatial patterns of diffusivity but the same approach can be to reveal other phenomena such as flows, scaffolding networks, and areas of trapping or confinement.

The examples analyzed in this work highlight the wide versatility of MAGIK. Remarkably, the same architecture can be applied to investigate other dynamical observables, can be trained to simultaneously estimate several parameters, and can even be used for applications beyond time-lapse microscopy, where time is substituted by another variable. As such, it will enable new experimental designs and high-throughput analysis to decipher biological mechanisms underlying the modulation of spatiotemporal behavior.

\hypertarget{sec:methods}{}

\section*{Methods}

\subsection*{Description of MAGIK.}

The input to MAGIK is the graph representation of the movement and interactions of an ensemble of objects. The nodes ($\mathcal{V}$) contain features encoding meaningful information of the objects, and the edges ($\mathcal{E}$) connect spatiotemporally-neighboring nodes codifying relational features, such as, e.g.,  the Euclidean distance between them (Fig.~\ref{fig:1}a-b). 

The architecture comprises three main blocks. First, an encoder neural network $\phi_v$ converts each node feature representation $v_i \in \mathcal{V}$ of dimension $l$ into a ${l}'$-dimensional feature representation ${v_i}'$ (Fig.~\ref{fig:1}c). In parallel, another encoder neural network function $\phi_e$ transforms each 1-D edge feature $e_k \in \mathcal{E}$ into a high-level feature vector ${e_k}'$ of dimension ${f}'$ (Fig.~\ref{fig:1}d). $\phi_v$ and $\phi_e$ are a series of multi-layer perceptrons (MLPs) composed of a linear layer followed by a Gaussian Error Linear Unit (GELU)~\cite{hendrycks2016gaussian} as activation function and layer normalization. 

Second, the resultant graph representation $\mathcal{G}~=~\left \{ {\mathcal{V}}', {\mathcal{E}}'\right \}$ ((Fig.~\ref{fig:1}e)) is processed through repeated fingerprinting graph blocks (FGNN, described in detail in Suppl. Fig.~\ref{fig:s1}a-f). Each FGNN updates each edge in the graph by applying an MLP to the concatenation of the features of two neighboring nodes and their connecting edge, i.e., 
\begin{equation}
    {e}^{''}_{ij} = \textup{MLP}\left ( \left [ {v_i}', {v_j}', {e}'_{ij}  \right ] \right ) \enspace,
    \label{eq:e_ij}
\end{equation}
for $j \in \mathcal{N}_i$, where $\mathcal{N}_i$ is the neighborhood of node $i$, and $\left [ , \right ]$ represents the concatenation operation (Fig.~\ref{fig:s1}b). Subsequently, the learned representation ${e}^{''}_{ij}$ (of dimension ${f}'$) is weighted by a Gaussian attention mechanism, 
\begin{equation}
w_{ij} = \exp\left ( - \left ( \frac{d_{ij}^{2}}{2 \sigma^{2}} \right )^{\beta} \right ) \enspace,
\label{eq:w_ij}
\end{equation}
where $d_{ij}$ is the Euclidean distance between the centroids of the nodes $i$ and $j$, and the standard deviation $\sigma$ and the Gaussian order $\beta$ are learnable parameters that allow the FGNN to adapt to varied object dynamics (Figs.~\ref{fig:s1}c and  \ref{fig:s1}d). The FGNN computes a local representation for the topological neighborhood $\mathcal{N}_i$ by applying a linear transformation to the concatenation of the current state of node $i$ and the aggregate of the weighted edge features, according to
\begin{equation}
h_i =  \mathbf{W}_H\left [{v}'_i, \sum_{j \in \mathcal{N}_i}w_{ij}{e}^{''}_{ij}  \right ] \in H \enspace,
\label{eq:h_i}
\end{equation}
where $\mathbf{W}_H$ is a ${l}' \times ({l}' + {f}')$ linear projection matrix. Importantly, we prepend a learnable node embedding $U~\in~\mathbb{R}^{{l}'}$ to the local representation matrix, i.e., $H~=~\left [ U; H \right ]$, whose state serves as a graph-level representation (Fig.~\ref{fig:s1}e)~\cite{dosovitskiy2020image}. Finally, gated self-attention layers~\cite{jumper2021highly} are used to update the hidden states of the node features,
\begin{equation}
\begin{aligned}
{\mathcal{V}}''^{(z)} & = \textup{attn}^{(z)}(H)  \\ & = \mathbf{G^{(z)}} \odot   \left ( \textup{softmax}\left ( \frac{1}{\sqrt{c}}\mathbf{Q^{(z)}}\mathbf{K^{(z)}}^{\top}
 \right )\mathbf{P^{(z)}} \right ) \enspace,
\end{aligned}
\label{eq:attn}
\end{equation}
where $z = 1, \cdots, Z$, with $Z$ representing the number of attention heads; $\mathbf{Q}^{(z)}~=~H\mathbf{W}_Q^{(z)}$, $\mathbf{K^{(z)}}~=~H\mathbf{W}_K^{(z)}$, and $\mathbf{P^{(z)}}~=~H\mathbf{W}_P^{(z)}$ are the queries, key, and values,  embedding matrices of dimension $c$ obtained by the ${l}' \times {l}'$ linear projection matrices $\mathbf{W}_Q^{(z)}$, $\mathbf{W}_K^{(z)}$, $\mathbf{W}_P^{(z)}$, respectively; $\mathbf{G}^{(z)}~=~\sigma\left ( H\mathbf{W}_G^{(z)} \right )$ is the gate vector parametrized by the linear projection matrix $\mathbf{W}_G^{(z)} \in \mathbb{R}^{{l}' \times {l}'}$, followed by an element-wise sigmoid function $\sigma$; $\odot$ denotes the Hadamard product; and softmax normalizes the self-attention weights to be positive and add up to 1. The multi-head outputs ${\mathcal{V}}''^{(z)}$ are concatenated and passed through a MLP to capture non-linear interactions between the node features to provide the updated node embbedings~${\mathcal{V}}''$ (Suppl. Fig.~\ref{fig:s1}f). Note that ${U}'$ needs to be retrieved from ${\mathcal{V}}''$ to obtain the updated global features.

Third, the final node (${\mathcal{V}}''$), edge (${\mathcal{E}}''$), and global features (${U}'$) are decoded to obtain node, edge, and global-level predictions. The node features ${\mathcal{V}''}$ are processed using the decoding neural network $\varphi_v$ to obtain predictions for nodes. Similarly, the decoder neural network $\varphi_e$ receives ${\mathcal{E}}''$ and yields a prediction for each edge in the graph. $\varphi_v$ and $\varphi_e$ are reflections of the encoder networks $\phi_v$ and $\phi_e$, respectively, with an additional (prediction) layer comprising a linear transformation tailed by an output activation function (e.g., softmax or logistic sigmoid for classification problems, or linear activation for regression tasks). To compute global attributes, $U'$ is processed by $\varphi_u$, an MLP followed by a linear layer and a task-dependent nonlinear activation.

To demonstrate the versatility of MAGIK, we use the same model architecture for all examples. The encoding neural networks $\phi_v$ and $\phi_e$ consist of a series of MLPs of dimension 32, 64, and 96, respectively. The latent dimension for nodes and edges (i.e., ${l}' = {f}' = 96$) is maintained across two FGNNs layers in the trunk of the network and is chosen such that it is divisible by the number of self-attention heads in each layer ($Z = 12$). The global embedding vector $U$ is zero-initialized. The node and edge decoding neural networks $\varphi_v$ and $\varphi_e$ consist of three MLPs of dimensions 96, 64, 32, followed by a final linear layer and an activation function that map the decoded node and edge features to the output dimension. $\varphi_u$ consists of a 64-dimensional MLP followed by a linear output layer and an activation function that returns the global-level predictions.

\subsection*{MAGIK training.}

Once the network architecture is defined, MAGIK is trained using a set of graph feature representations and task-dependent targets. The input graphs follow the same relational structure regardless of the task, with nodes describing object detections and edges connecting the objects in time and space. Targets, in turn, represent different parameters depending on the specific task. 

For {\it trajectory linking} (Figs.~\ref{fig:2}-\ref{fig:3}),  MAGIK is trained to predict the probability of having a connection/link between two objects. This task is modeled as an edge-classification problem with a binary label (linked, labeled with 1, or unlinked, labeled with 0). Thus, during training, the network aims at minimizing the binary cross-entropy between the predicted probabilities and the ground-truth label for each edge. Accordingly, $\varphi_e$ uses a sigmoid function as the final activation to produce probability estimates. 
For the trajectory linking tasks, MAGIK processes input graphs in batches of 8 samples while training. Each sample is obtained from a fraction of frames (10 to 20\%), stochastically extracted from the same training video. Graphs are created according to Fig.~\ref{fig:1}a and augmented by translations, rotations, and mirroring of the set of nodes' centroids. Likewise, the object descriptors are augmented by adding random noise to their values.  Moreover, we randomly remove nodes and their connections to account for detection blinking. For all the trajectory linking examples, the network was trained for 100 epochs, each consisting of 512 unique training samples split into batches of 8. 

The {\it inference of local properties} is modeled as a node-regression problem (Figs.~\ref{fig:4}a-d), where MAGIK trains to minimize the mean absolute error (MAE) between node predictions and ground truth. Here, $\varphi_v$ uses a linear activation function as the output activation. The training data comes from 2000 videos simulated with heterogeneous sets of moving objects and varying lengths (between 50 and 55 frames), and  their graph representations are augmented by translations, rotations, and mirroring of the nodes' centroids (further details are provided in the ``Simulations'' section). As ground truth, we used either the diffusion coefficient (Figs.~\ref{fig:4}b-d) or the anomalous diffusion exponent (Figs.~\ref{fig:s2}b-d) of the object at the node level. In these examples, the network was trained for 100 epochs, each consisting of 1024 unique training samples split into batches of 8.

The {\it quantification of global dynamic properties} requires MAGIK to be trained to estimate global-level attributes from the input graphs. Throughout the examples, we have approached this problem from different perspectives,  from a classification problem to determine the underlying diffusion model of a set of particles (Figs.~\ref{fig:4}e-l) to a regression problem to estimate the relative fraction of objects moving according to different diffusion modes (Suppl. Figs.~\ref{fig:s2}e-h). For classification tasks, the network is trained to minimize the sparse categorical cross-entropy between class predictions and ground truth labels, with a softmax as the output activation of $\varphi_u$. For regression tasks, MAGIK minimizes the MAE between the network estimates and the target features. Here, $\varphi_u$ uses a linear activation function as the output activation. In each of these examples, the training data comes from 2000 simulated videos from which we extracted graph representations and augmented their topological structure by translations, rotations, and mirroring of the nodes' centroids. As target features, we used either class labels (for classification tasks) or continuous features (for regression tasks). The network was trained for 100 epochs, each consisting of 1024 unique training samples split into batches of 8.

For all examples, the trainable parameters of MAGIK (i.e., the weights of the artificial neurons in the neural networks and the parameters of Gaussian edge weighting function) were iteratively optimized using the backpropagation training algorithm~\cite{mcclelland1987parallel} and Adam optimizer (with a learning rate of 0.001)~\cite{kingma2014adam}. Furthermore, new training data was continuously generated during training. The training time of MAGIK ranges between 1 and 5 mins for trajectory linking and from 30 to 60 mins in the case of node and global-level regression, on an NVIDIA A100 GPU (40 GB VRAM, 2430 MHz effective core clock, 6912 CUDA cores). 

\subsection*{Postprocessing algorithm for trajectory linking.}

Cell trajectories are built from the scores obtained for the edge classification problem through a simple postprocessing algorithm.
The algorithm starts from a random node at the initial frame $t=0$ and connects it over time with other nodes at subsequent frames, considering only edges that have been classified as ``linked'' by MAGIK. If no ``linked'' edges connect the sender node at $t$ with any receiver nodes at $t+1$, the algorithm checks future frames, until a maximum time lag. If no ``linked'' edges are found within this lag, the trajectory is interrupted. If a sender node has two ``linked'' edges connecting it to two receiver nodes at a later frame, the event is identified as a division. At this point, the algorithm treats the two nodes as independent and attempts to build two new trajectories. In the rare event that more than two ``linked'' edges originate from the same sender, the one connecting the furthest receiver is dropped. The procedure is iterated until all the ``linked'' edges have been taken into account.

\subsection*{Quantification of cell tracking results.}

Quantification of the method performance for cell tracking was obtained by calculating the $\operatorname{TRA}$ metric based on the acyclic oriented graph matching ($\operatorname{AOGM}$) measure discussed in Ref.~\cite{matula2015cell}.
First, images corresponding to the incomplete cell segmentation provided for the $6^{\rm th}$ Cell Tracking Challenge were annotated according to their ground truth and then transformed into an acyclic oriented graph according to the instructions for participation in the challenge~\cite{ulman2017objective}. A similar graph was also obtained for the trajectories predicted by our methods. The quantification of the matching between the two graphs performed by the AOGM corresponds to the weighted sum of the executed operations to transform the predicted graph into the ground-truth one~\cite{matula2015cell}. For this, we used the $\operatorname{AOGM-A}$ measure, which corresponds to the $\operatorname{AOGM}$ measure calculated by keeping only the edge-related weights positive ($w_{\rm NS} = w_{\rm FN} = w_{\rm FP} = 0$; $w_{\rm ED}=1$, $w_{\rm EA}=1.5$, $w_{\rm EC} =1$)~\cite{matula2015cell}. The $\operatorname{AOGM-A}$ thus evaluates the ability of an algorithm to follow objects in time (i.e., its linking capability). The $\operatorname{AOGM-A}$ measure is normalized to obtain the tracking accuracy ($\operatorname{TRA}$):
\begin{equation}
\operatorname{TRA} = 1-\frac{{\rm min}({\operatorname{AOGM-A},\operatorname{AOGM-A}_0)}}{\operatorname{AOGM-A}_0},  
\label{TRA}
\end{equation}
where $\operatorname{AOGM-A}_0$ corresponds to the cost of linking the graph from scratch (i.e., the cost of adding all the edges multiplied by the corresponding weights). The normalization bounds ${\rm TRA}$ in the interval $[0,1]$, with higher values corresponding to better tracking performance.

\subsection*{Simulations}
Trajectories were simulated using the {\tt andi-datasets} Python package~\cite{andigithub}. Additionally, we used DeepTrack $2.1$ to render imaged objects in different illumination modalities (fluorescence and holographic microscopy) reproducing  optical conditions to provide realistic node features (Fig.~\ref{fig:4}).

For the fluorescence microscopy experiments of Fig.~\ref{fig:4}a-d and Suppl. Fig.~\ref{fig:s2}a-d, we simulated objects performing FBM in two dimensions with random anomalous exponents ($0.2 \leq \alpha < 1.8$) and diffusivities ($0.005 \leq D < 0.7$). For Fig.~\ref{fig:4}e-h, the diffusivity was defined by a random spatial map, smoothed with a Gaussian filter. For training, we typically use videos of $50-55$ frames containing $30-35$ objects ($70-80$ for varying diffusivity and diffusivity maps) initially positioned at random locations. Each object is rendered as a diffraction-limited spot through the optics module of DeepTrack 2.1, with a random intensity from a uniform distribution between 20 and 80 counts, varying over time with a standard deviation of $3$ counts.

For all the experiments of Fig.~\ref{fig:5}, we generated trajectories undergoing three different diffusion models, namely FBM, ATTM, CTRW, with a constant anomalous exponent $\alpha = 1$ and random diffusivity. For Fig.~\ref{fig:5}a-e, each object in the video undergoes 2D diffusion with a randomly-assigned  model, with all other properties (sequence length, number of particles, intensity) being the same as described for Fig.~\ref{fig:4}.

For the plankton trajectories illustrated in Fig.~\ref{fig:5}f-m, all microorganisms in the same video move according to the same 3D model, varying from video to video. We generate holographic videos of $100$ frames including $3-7$ microorganisms, each with a randomly sampled refractive index from a uniform distribution between $1.35 - 1.55$, covering a wide variety of plankton species in the literature \cite{aas1996refractive}.

For the examples illustrated in Suppl. Fig.~\ref{fig:s2}e-h, we generate fluorescence images of objects undergoing FBM in two dimensions in sub-diffusive ($0.2 \leq  \alpha \leq  0.6$), normal ($\alpha=1$), and super-diffusive mode ($1.4 \leq  \alpha \leq  1.8$). All other properties (sequence length, number of particles, intensity) are the same as described for Fig.~\ref{fig:4}.

\section*{Data Availability}
The cell tracking datasets were obtained from the cell tracking challenge webpage \url{http://celltrackingchallenge.net/2d-datasets/}, where they can be accessed from. 

\section*{Code Availability}
All source code and examples are made publicly available at the DeepTrack-2.1 GitHub repository \cite{deeptrackgithub}.

\bibliography{biblio}

\section*{Acknowledgments}
The authors thank Gorka Mu{\~n}oz-Gil, Henrik Klein, and Fredrik Sk{\"a}rberg for useful discussions.

\section*{Funding}
JP, BM, HB, and GV,  were supported by the H2020 European Research Council (ERC) Starting Grant ComplexSwimmers (Grant No. 677511), the Horizon Europe ERC Consolidator Grant MAPEI (Grant No. 101001267), the Knut and Alice Wallenberg Foundation (Grant No. 2019.0079).
CM was supported by funding from FEDER/Ministerio de Ciencia, Innovaci\'{o}n y Universidades -- Agencia Estatal de Investigaci\'{o}n through the ``Ram\'{o}n y Cajal'' program 2015 (Grant No. RYC-2015-17896), and the ``Programa Estatal de I+D+i Orientada a los Retos de la Sociedad'' (Grant No. BFU2017-85693-R); from the Generalitat de Catalunya (AGAUR Grant No. 2017SGR940). 
CM also acknowledges the support of NVIDIA Corporation with the donation of the Titan Xp GPU.

\section*{Contributions}
CM and GV conceived the project. 
JP and CM designed the method. 
JP, BM, and SN implemented the architecture. 
JP, HB, BM, and CM analyzed the data and generated the figures. 
HB, JP, and CM carried out the simulations.
JP, GV, and CM wrote the paper with input from all of the authors. 
DM, GV, and CM supervised the project.

\section*{Corresponding authors}
Correspondence should be addressed to \href{mailto:giovanni.volpe@physics.gu.se}{Giovanni Volpe} or \href{mailto:carlo.manzo@uvic.cat}{Carlo Manzo}.

\section*{Competing interests}
The authors declare no competing interests.

\appendix
\clearpage
\onecolumngrid
\section*{Supplementary Material}

\makeatletter
\renewcommand{\figurename}{SUPPL. FIG.}
\setcounter{figure}{0}
\setcounter{page}{1}
\makeatother
   
\subsection*{Supplementary Figures}

\begin{figure*}[htb!]
    \makeatletter 
    \renewcommand{\thefigure}{S\@arabic\c@figure}
    \renewcommand{\theHfigure}{Supplement.\thefigure}
    \makeatother
    \centering
    \includegraphics[width=0.9\textwidth]{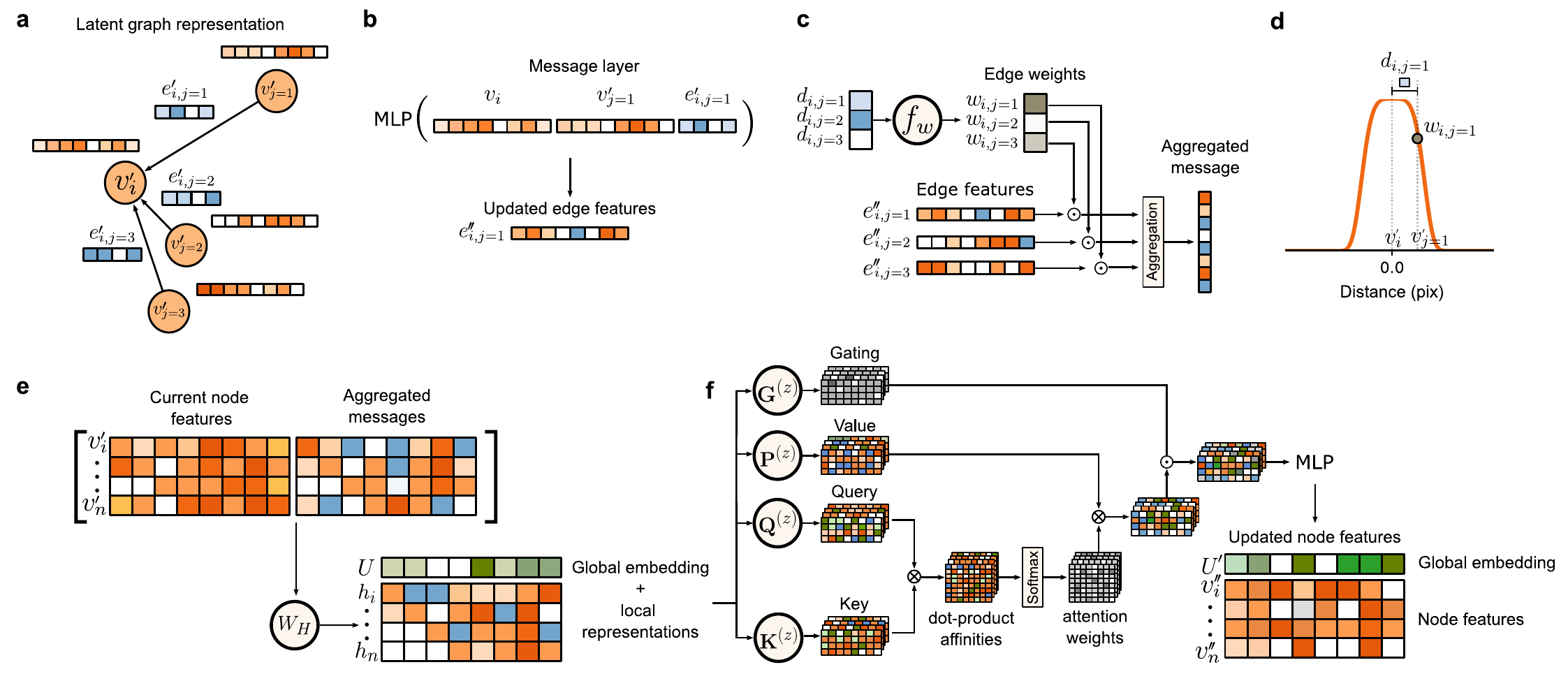}
     \caption{
     \textbf{Processing flow of the fingerprinting graph block (FGNN)}.  
     FGNN, similar to other flavors of GNN layers, comprises three fundamental steps: edge feature update, edge feature aggregation, and node update.
     \textbf{a}, Input graph structure. Nodes contain features encoding the object's position and relevant descriptors. Edges encode relational features between neighboring nodes.
     In this example, the node of interest, labeled with the subindex $i$, receives information from connected nodes, labeled with the subindex $j$.
     \textbf{b}, Each edge in the graph is updated by applying a multilayer perceptron (MLP) to the concatenation of the features of two nodes and the edge connecting them (Eq.~\eqref{eq:e_ij}).  \textbf{c}, During the aggregation of edge features to a node, the contribution of each edge has a weight that is determined by the distance between linked nodes using a function with free parameters, $f_w$ (Eq.~\eqref{eq:w_ij}). 
     \textbf{d} $f w$ is a super-Gaussian and defines a learnable local receptive field that allows the FGNN to adapt to heterogeneous dynamics. 
     \textbf{e}, The current state of the nodes and the aggregate of the weighted edge features are concatenated and linearly transformed to obtain a local representation for each neighborhood (Eq.~\eqref{eq:h_i}). Furthermore, the FGNN prepends a learnable node embedding $U$ to the local representation matrix, whose features provide global system-level insights. 
     \textbf{f} The nodes are updated using gated self-attention layers. The matrix resulting from the concatenation of $U$ with the local features is transformed by the trainable linear transformation matrices $\mathbf{Q}^{(z)}$, $\mathbf{K}^{(z)}$, $\mathbf{P}^{(z)}$ to obtain queries, key, and values, respectively. $z$ denotes the index of the attention head. The self-attention weights are calculated by the dot-product of the queries with the key matrix. Softmax normalizes the weights to be positive and to add up to 1 (Eq.~\eqref{eq:attn}). Finally, the weighted values are multiplied by the gatings and passed through an MLP to account for non-linear interactions between nodes to obtain the updated node features.
     }
    \label{fig:s1}
\end{figure*}

\begin{figure*}[htb!]
    \makeatletter 
    \renewcommand{\thefigure}{S\@arabic\c@figure}
    \renewcommand{\theHfigure}{Supplement.\thefigure}
    \makeatother
    \centering
    \includegraphics[width=\textwidth]{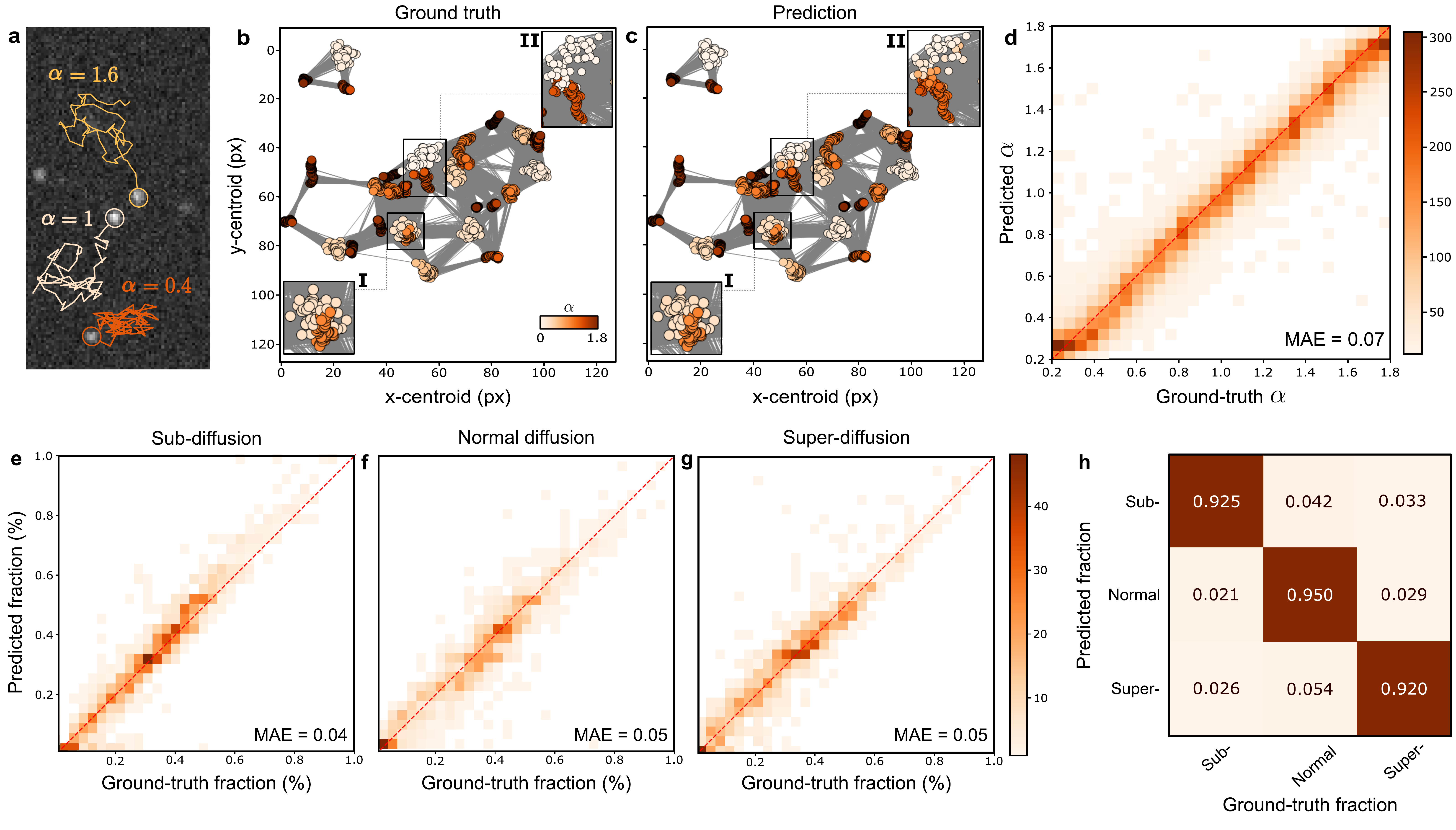}
     \caption{
     \textbf{MAGIK estimates local and global anomalous diffusion properties at the ensemble and single-object levels.} 
    \textbf{a}, Simulated single-object tracking experiment. Fluorescence microscopy is used to follow the motion of single molecules characterized by a fractional Brownian motion (FBM) with varying anomalous diffusion exponent $\alpha$. 
    \textbf{b, c}, Ground-truth and predicted graphs. Edges depict the network of associations used to directly infer dynamic properties without explicit linking. Nodes are color-coded according to the value of the target feature $\alpha$. The predicted node values agree with the ground-truth also in crowded areas (e.g., zoomed regions I and II). 
    \textbf{d}, Probability distribution of the predicted vs. ground-truth anomalous diffusion exponent $\alpha$.
    \textbf{e-h}, MAGIK estimates the relative fraction of objects following different diffusion modes, i.e., sub- ($0.2 \leq  \alpha \leq  0.6$), normal ($\alpha = 1$) and super-diffusion  ($1.4 \leq  \alpha \leq  1.8$).
    \textbf{e-g}, Probability distribution of predicted vs. ground-truth fraction for sub-diffusion, normal diffusion, and super-diffusion, respectively.
    \textbf{h} Confusion matrix demonstrating how the network classifies the underlying diffusion model exhibited by objects in 1199 validation videos. Column-based normalization is applied, such as the sum along the columns adds up to 1, with minor deviations due to rounding.
    }
    \label{fig:s2}
\end{figure*}

\clearpage
\subsection*{Supplementary Videos}
    \makeatletter
    \renewcommand{\figurename}{SUPPLEMENTARY VIDEO}

\setcounter{figure}{0}
\begin{figure*}[htb!]

    \renewcommand{\thefigure}{S\@arabic\c@figure}
    \renewcommand{\theHfigure}{Supplement.\thefigure}
    \makeatother
    \centering
     \caption{\textbf{Linking of HeLa cells.}
     MAGIK successfully tracks ($\operatorname{TRA}=99.2\%$) HeLa cells on a flat glass substrate despite being shape-varying cells with high packing density, low SNR, and heterogeneous dynamics as a consequence of their migration and proliferation. Cells are color-coded according to their trajectory index.}
    \label{vid:s1}
\end{figure*}

\begin{figure*}[htb!]
    \makeatletter
    \renewcommand{\thefigure}{S\@arabic\c@figure}
    \renewcommand{\theHfigure}{Supplement.\thefigure}
    \makeatother
    \centering
     \caption{\textbf{Linking of GFP-GOWT1 cells.} 
     MAGIK successfully tracks GFP-GOWT1 mouse stem cells even as they leave the field of observation ($\operatorname{TRA}=99.2\%$). Cells are color-coded according to their trajectory index.}
    \label{vid:s2}
\end{figure*}

\begin{figure*}[htb!]
    \makeatletter
    \renewcommand{\thefigure}{S\@arabic\c@figure}
    \renewcommand{\theHfigure}{Supplement.\thefigure}
    \makeatother
    \centering
     \caption{
     \textbf{Linking of U373 cells.} 
     MAGIK accurately tracks glioblastoma-astrocytoma U373 cells on a polyacrylamide substrate despite these being shape-varying cells ($\operatorname{TRA}=100\%$). Cells are color-coded according to their trajectory index.}
    \label{vid:s3}
\end{figure*}

\begin{figure*}[htb!]
    \makeatletter
    \renewcommand{\thefigure}{S\@arabic\c@figure}
    \renewcommand{\theHfigure}{Supplement.\thefigure}
    \makeatother
    \centering
     \caption{\textbf{Linking of fluorescent HeLa cells.} 
     MAGIK accurately tracks HeLa cells stably expressing H2b-GFP despite the dense sample and the frequent mitosis and cell collisions ($\operatorname{TRA}=98.4\%$). Cells are color-coded according to their trajectory index.}
    \label{vid:s4}
\end{figure*}

\begin{figure*}[htb!]
    \makeatletter
    \renewcommand{\thefigure}{S\@arabic\c@figure}
    \renewcommand{\theHfigure}{Supplement.\thefigure}
    \makeatother
    \centering
     \caption{\textbf{Linking of pancreatic stem cells.} 
     MAGIK successfully tracks pancreatic stem cells on a polystyrene substrate despite high cell density, elongated shapes, pronounced cell displacements, and a significant number of division events ($\operatorname{TRA}=98.5\%$). Cells are color-coded according to their trajectory index.}
    \label{vid:s5}
\end{figure*}

\end{document}